\documentclass[psfig]{aa}
\usepackage{epsfig}

\def\apgt{\ {\raise-.5ex\hbox{$\buildrel>\over\sim$}}\ }
\def\aplt{\ {\raise-.5ex\hbox{$\buildrel<\over\sim$}}\ }

\newcommand{\kms}{\mbox{${\rm km~s}^{-1}$}}

\begin{document}

\offprints{Q.Z. Liu (email: qzliu@astro.uva.nl)}
\thesaurus{}

\title{New emission phase of the Be/X-ray binary X Persei: 
nearly simultaneous spectroscopic and near-infrared observations}
\author{Q.Z. Liu\inst{1, 2}, 
	H.R. Hang\inst{1}}
\institute{Purple Mountain Observatory, Nanjing 210008, China: National Astronomical Observatories, 
                      Chinese Academy of Sciences           
\and
           Astronomical Institute "Anton Pannekoek", University of Amsterdam, Kruislaan 403, 
           1098 SJ Amsterdam, The Netherlands}

\date{Received 4 December 1999 / Accepted 15 August 2000}
\maketitle
\markboth{Q.Z.~Liu \& H.R.~Hang:\,\, New emission phase of the Be/X-ray binary X Per}{}

\begin{abstract}
We present a series of optical spectroscopic and near-infrared photometric observations of the 
Be star X Persei, from the beginning of the recent emission phase. Our data show that after the latest 
extended low state ($ELS$), both the equivalent width ($EW$) of the H$\alpha$ emission line and the $JHK$ 
luminosities gradually increased. The recent maximum H$\alpha$ $EW$ and $JHK$ magnitudes are more than but 
comparable to the 1986-1989 maxima, which reflects a more extensive and denser envelope in the new 
emission phase. 

The IR photometry and the H$\alpha$ $EW$ increase coincidently in the early stage of the new emission 
phase. However, the variations in the following years, especially in 1994 and 1995, are much different 
from the early variations. An unusual variation that while the H$\alpha$ $EW$ in X Per increases considerably, the 
$JHK$ magnitudes decline rapidly is seen in our nearly simultaneous spectroscopic and near-infrared 
observations.

An expanding ring model is put forward to interpret the observed unusual variations between the 
H$\alpha$ $EW$ and near-IR luminosity during the recent emission phase of X Persei. Our results indicate that 
although both infrared excess and H$\alpha$ emission line arise from the envelope, their main contributors, in 
some cases, are likely to originate in the different part of the envelope.

\end{abstract}

\section{Introduction}

X Persei, the optical counterpart of the X-ray source 4U0352+30 (Braes \& Miley 1972), is a sixth 
magnitude B0 Ve star (Lyubimkov et al. 1997; Roche et al. 1997) with a magnetized neutron star 
companion (pulse period ~835s). Possible orbital period of the system is 580 days, detected in the 
radial velocities of the Balmer lines (Hutchings 1977). This period has not yet been confirmed, and 
is even doubted (Penrod \& Vogt 1985; Reynolds et al. 1992). The star is also a known optical 
variable on time scales from minutes to years.  

X Persei sometimes showed extreme, extended low states (Mook et al. 1974; Roche et al. 1993). 
Optical and near infrared photometry indicated that X Persei dropped to an extreme low state begun in 
1989 March (Norton et al. 1991; Roche et al. 1993). During this period the H$\alpha$ line profile changed from 
emission to absorption, accompanied by a decrease in the IR flux by over a magnitude (Norton et al. 1991; 
Reynolds et al. 1992; Roche et al. 1993). This behavior indicated the loss of the circumstellar disk around 
the Be star as is also known from single Be star. This opportunity has been used for determining the 
physical parameters of the primary star (Fabregat et al. 1992; Lyubimkov et al. 1997). Corbet \& Thomas (1991) 
reported the 
H$\alpha$ line in emission again in October 1991, which meant that a new circumstellar disk began to form. 
More recently, Zamanov \& Zamanova (1995) found the optical low state at $V$ band begun in the mid-1990 
had ended in the spring of 1993. The system reached maximum brightness in late 1994, and has 
subsequently entered a period of relatively rapid fading. The most comprehensive optical and IR 
variations of X Per over the past decades, covered the latest $ELS$, were presented by Roche et al. (1993; 
1997) and Telting et al. (1998). The spectral observations in Feb. 1995 showed the HeI l 6678 was in 
emission with quadruple peaks, implying the existence of a double disc structure (Tarasov \& Roche 1995; 
Kunjaya \& Hirata 1995). 

In this paper, we present the nearly simultaneous optical spectroscopy and near-IR 
photometry of this interesting object from the beginning of the new emission phase. It is a good 
opportunity to trace the build-up of a new envelope and the subsequent evolution of the new envelope in X 
Persei. Some early observations were reported in a previous paper (Liu \& Hang 1997). Finally, we will 
put forward a model to give an explanation of the observed phenomenon between the H$\alpha$ $EW$ and near-IR 
luminosity.

\section{Observations and results}

Since the autumn of 1992, we have chosen a set of Be stars, especially Be/X-ray binaries, for spectroscopic 
and near-IR luminosity monitoring. X Persei was one of the program stars. All observations were carried 
out at Xinglong station of Beijing Astronomical Observatory and at Yunnan Astronomical Observatory. 

\subsection{Near-IR photometry}

The near-IR photometry was measured at Xinglong station of Beijing Astronomical Observatory by using 
the 1.26 m reflectors with a liquid nitrogen cooled InSb photometer covering the $JHK$ bands. Typical 
errors are $\pm$ 0.05 for $J$, $H$ bands, and $\pm$ 0.07 for $K$ band. The results of measurements are presented in 
Table I.

\begin{table}[t]
\caption{ Near-IR measurements of X Persei/4U0352+30.}
\begin{tabular}{crccc}
\noalign{\smallskip}
\hline
\noalign{\smallskip}
  Date      & Julian date & J                          & H                           & K                           \\
                 &  (-2440000)   & (1.25$\mu$m)   &  (1.25$\mu$m)    &  (1.25$\mu$m)   \\ 
\noalign{\smallskip} 
\hline
\noalign{\smallskip}
 92.11.11 &  8938.18     & 6.51  &  6.54  &    6.51     \\
 92.11.18 &  8945.33     & 6.56  &  6.30  &    6.54     \\
 93.11.07 &  9299.26     & 6.01  &  5.71  &    5.44     \\
 94.09.17 &  9614.29     & 5.56  &  5.40  &    5.10     \\
 94.09.19 &  9616.30     & 5.32  &  5.26  &    5.19     \\
 94.09.20 &  9617.34     & 5.60  &  5.47  &    5.32     \\
 94.09.22 &  9619.31     & 5.45  &  5.32  &    5.26     \\
 95.11.07 & 10029.33    & 6.23  &  6.23  &    6.08     \\
 95.11.11 & 10033.28    & 6.17  &  6.15  &    6.01     \\  
\noalign{\smallskip}
\hline
\end{tabular}
\end{table}

Fig. 1 shows the observed IR lightcurves of X Persei in the $JHK$ bands, together with the results 
from Norton et al. (1991), Reynolds et al. (1992), Roche et al. (1993), and Telting et al. (1998). For 
comparison we also collect the published data of the optical $V$ magnitude during the same period 
(Zamanov \& Zamanova 1995) and include them in the figure. One can see that the brightness in $JHK$ 
was almost constant during the $ELS$, at a level of $J$, $H$ and $K$ approximate to $6.6$ magnitude. The $ELS$ 
ended with a gradual increase in brightness, reaching a flat high state peaked around JD 2449700, in 
agreement with the optical data. The peak $V$ magnitude recovered its 1986-1989 maximum brightness of 
6.2$^m$, with an increase of $\bigtriangleup$$V$$\approx$ 0.6$^m$ over the magnitude at the latest $ELS$. 
All the lightcurves show a similar pattern, and the IR brightness variations follow those of the V-band 
lightcurve, but with greater amplitude: $\bigtriangleup$J$\approx$ 1.3$^m$, $\bigtriangleup$$H$$\approx$ 1.4$^m$, 
$\bigtriangleup$$K$$\approx$ 1.5$^m$. This recent maximum is of comparable brightness, in 
all IR bands, to the 1986-1989 maximum. The observations in October 1995, however, showed a rapid fall 
to $J=6.2^m$, $H=6.1^m$, $K=6.0^m$, close to the magnitudes at the $ELS$.

\begin{figure}[ht]
\vspace{0cm}
\hspace{-2.04cm} 
\psfig{file=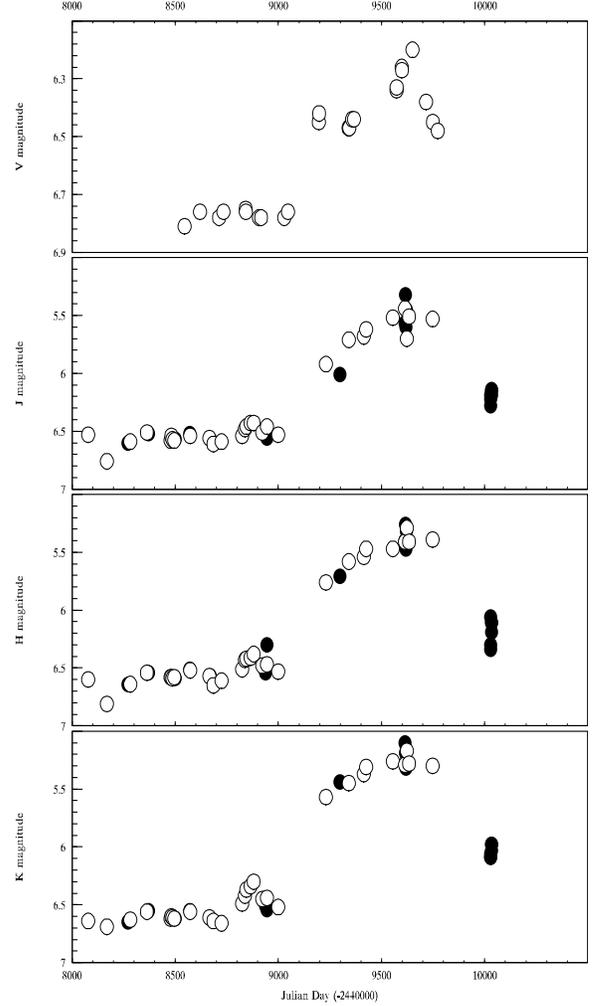, height=14cm, width=14cm, angle=0}
\caption{The near-IR $JHK$ magnitudes during the course of the monitoring programme. Open circles are 
from Roche et al. (1993), Norton et al. (1991), Reynolds et al. (1992), and Telting et al. (1998); filled 
circles are from this work. }
\end{figure}

\subsection{Optical spectroscopy}

The optical spectroscopic observations were mainly made at Xinglong station of Beijing Astronomical 
Observatory with a CCD grating spectrograph at the Cassegrain focus of the 2.16 m telescope. Two 
additional high-resolution spectra were performed at the Coude focus of the 100cm telescope at Yunnan 
Astronomical Observatory. The journal of the observations for X Per is summarized in Table II, together 
with the $EW$ of H$\alpha$ and HeI 6678 lines. Sometimes the object was observed several times at one night. He-
Ar spectra were taken in order to obtain the pixel-wavelength relations. All spectroscopic data are reduced 
with IRAF software on our Sun-4 station except the spectra taken in 1992, which were reduced with 
FIGARO/DIPSO software on the Vax 11/780 computer of Purple Mountain Observatory. They are bias 
subtracted and flat field corrected. The influence of night sky light on the spectra has been eliminated.

\begin{table*}[t]
\caption{ Journal of the spectroscopic observations of X Persei/4U0352+30.}
\begin{tabular}{crrcrr}
\noalign{\smallskip}
\hline
\noalign{\smallskip}
  Date      & Julian date & Dispersion &  Resolution  & H$\alpha$ $EW$ & HeI 6678 $EW$  \\
                 &  (-2440000)   & (\AA /mm)    &  (\AA /pixel)  &   (\AA)                 & (\AA)                 \\ 
\noalign{\smallskip} 
\hline
\noalign{\smallskip}
 92.11.01 &  ~8928.2      & 101   &  2.26     & -0.93    &   0.18      \\
 92.11.03 &  ~8930.1      & 101   &  2.26     & -1.39    &   0.16      \\
 92.11.04 &  ~8931.2      & 101   &  2.26     & -1.45    &   0.16      \\
 93.11.04 &  ~9296.2      &   50   &  1.39     &  -6.41    &   -0.29    \\
 93.11.07 &  ~9299.1      &   50   &  1.39     &  -6.35    &   -0.23    \\
 94.09.21 &  ~9618.2      &   50   &  1.22     &  -6.68    &   0.07      \\
 95.01.11 &   9729.1      &    - -   & 0.0478  & -10.45   &  - - -        \\
 95.10.15 & 10006.3      &   50   &  1.22     & -16.02   &  -1.59     \\
 95.10.17 & 10008.3      &   50   &  1.22     & -16.34   &   -1.79    \\
 96.10.25 & 10382.3      &   50   &  1.22     & -12.43   &   -1.00    \\
 96.10.26 & 10383.2      &   50   &  1.22     & -12.59   &   -1.07    \\
 97.12.15 & 10798.2      &   50   &  1.22     & -10.60   &   -0.78    \\  
 97.12.16 & 10799.2      &   50   &  1.22     & -10.92   &   -0.92    \\
 97.12.18 & 10801.2      &   50   &  1.22     & -11.58   &   -1.09    \\
 98.10.31 & 11118.2      &   50   &  1.22     & -9.706   &   -0.60    \\
\noalign{\smallskip}
\hline
\end{tabular}
\end{table*}

Spectroscopic observations of X Per show that the emission line profiles are highly variable. They are characterized 
by strong emission features of H Balmer and HeI lines. Fig. 2 shows the development of the emission-line profiles for 
the H$\alpha$ and HeI $\lambda$6678 between 1992 and 1998. All the profiles have been normalized to neighboring 
continuum. We respectively show several representative H$\alpha$ and HeI $\lambda$6678 line spectra from our total 
data set. Three of respectively H$\alpha$ and HeI $\lambda$6678 spectra were taken during the rise to maximum 
brightness (which occurred around 1994 November), and five H$\alpha$ and four HeI $\lambda$6678 
spectra during the subsequent fade. The line measurements of H$\alpha$ and HeI $\lambda$6678 for the double-peaked 
profiles are summarized in Table III, with intensity above continuum ($I$), peak separation ($\bigtriangleup$$v_p$) 
between $V$ 
and $R$ components, and the outer envelope radius ($R_s$) deduced from $\bigtriangleup$v$_p$ (Huang 1972) listed. 

\begin{figure}[ht]
\vspace{0cm}
\hspace{0cm} 
\psfig{file=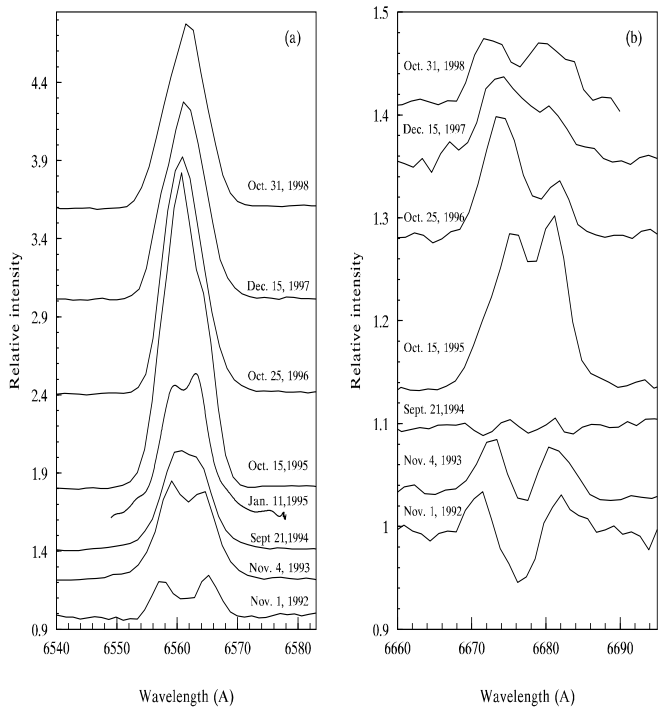, height=10cm, width=9cm, angle=0}
\caption{Series of the H$\alpha$ spectra (a) and the HeI $\lambda$6678 spectra (b) of X Persei during 1992-1998. All 
spectra have had the continuum level normalized and are annually offset vertically to allow direct 
comparison. Note the asymmetric H$\alpha$ profile in Oct. 1995, the significant decrease of the 
HeI $\lambda$6678 line intensity in 1994 and the asymmetric profile in 1996 and 1997.} 
\end{figure}

\begin{table*}[t]
\caption{Main parameters of the selected spectra of the H$\alpha$ and HeI $\lambda$6678 lines of X Persei.}
\begin{tabular}{rrcccccc}
\noalign{\smallskip}
\hline
\noalign{\smallskip}
Date & Julian date & \multicolumn{4}{c}{H$\alpha$}                                                           & \multicolumn{2}{c}{HeI 6678}             \\
          &  (-2440000)   & $I_V$  & $I_R$  & $\bigtriangleup$$v_p$(\kms) & $R_s (R_*)$ & $\bigtriangleup$$v_p$(\kms) &
 $R_s (R_*)$ \\
\noalign{\smallskip} 
\hline
\noalign{\smallskip}
 92.10.30 &  8926.3   & 1.20  &  1.25 & 378. & 1.12  & 477. & 0.70   \\
 93.11.07 &  9297.3   & 1.65  &  1.61 & 254. & 2.48  & 312. & 1.64   \\
 94.09.21 &  9618.3   & 1.66  &  1.66 & 167. & 5.74  & 274. & 2.13   \\
 95.01.11 &  9729.0   & 1.86  &  1.86 &  158. & 6.41 & - - -   & - - -     \\
 95.10.15 & 10006.3  &          &  3.03  & single peak   &  &  273. & 2.15 \\
 96.10.25 & 10382.3  &          &  2.52  & single peak   &  &  384. & 1.09  \\
 97.12.15 & 10799.3  &          &  2.27  & single peak   &  &  274. & 2.13 \\
 98.10.31 & 11118.3  &          &  2.17  & single peak   &  &  330. & 1.47 \\
\noalign{\smallskip}
\hline
\end{tabular}
\end{table*}

The H$\alpha$ spectra (Fig. 2a) during the first three years showed a well-defined double peak with that 
the intensity of $V$ component is nearly equal to that of $R$ component. The intensity enhancement was 
concentrated on the central parts of the line between two emission peaks with the peak separation 
gradually reduced. Such variations can be explained by the gradual expansion of the new envelope away 
from X Persei and/or the dissipation of the envelope (Hang \& Xia 1995; Liu \& Hang 1997). The line did 
not further increase and was still weak at the peak JHK magnitudes in 1994, but began to increase again 
in early 1995 with an asymmetric profile ($I_V$ / $I_R$$\approx$ 0.9). The line intensity reached maximum in October 
1995, and then gradually decreased over a few years with a single peak profile. The $V$ and $R$ components 
initially separated by $\sim$ 378 \kms~in the 1992 spectrum, but narrowing to $\sim$ 168 \kms~in early 1995. 

Fig. 2b shows that the HeI $\lambda$6678 line, after increasing during the first two years of the recent 
emission phase, almost disappeared in October 1994, with only a faint double peak ($EW=0.07$ \AA) superposed 
on the photospheric absorption line. In October 1995, however, there were large increases in both the blue 
and red peaks, as well as in the wings. While the line splitting in the H$\alpha$ had almost disappeared and a 
strong asymmetry was visible, the HeI $\lambda$6678 still remained a well-defined double peak. The four peaks in 
1995 February spectra reported by Tarasov \& Roche (1995), however, did not find in our 1995 October 
data. The features are narrow, so maybe the medium resolution spectra thta we have simply do not resolve the 
multiple peaks (Roche, private communication). The HeI line in 1992 spectrum had broad $V$ and $R$ 
wings, with a separation of $\sim $ 477 \kms. This evolved to a narrower line profile ($v_p\sim $ 274 \kms) in 
1994. We note that the shape of the He I line throughout these years is always double-peaked, and the 
two peaks differ very little in intensity, exception for the 1996 and 1997 spectra. 

The H$\alpha$ $EW$ varies more gradually than $V$ and $JHK$ magnitudes. Fig. 3 shows the H$\alpha$ $EW$ (lower 
panel) and $J$ magnitude (upper panel) as a function of time. Some H$\alpha$ $EW$ data are from Roche et al. 
(1993) and Kunjaya \& Hirata (1995). From the accumulated data one can see that the increase of the H$\alpha$ 
$EW$ started earlier than the brightening in the near-infrared $J$ magnitude. It began to increase at JD 
2448600 while the optical $V$ and near-infrared $JHK$ magnitudes were still at extremely low level. The 
dramatically increasing $EW$ is immediately apparent with the H$\alpha$ $EW$ changing from -0.2 \AA~ in 
1992 to -16.3 \AA~ in 1995. It is interesting to note that the $EW$ in 1994 did not increase to its peak, as the $J$ 
magnitude did. Its $EW$ was only -6.5 \AA~ at the peak $J$ magnitude, not much different from its value in 1993. 
The H$\alpha$ $EW$, however, was seen to dramatically increase during the optical and near-infrared fade. While 
the latter got fainter dramatically, the $EW$ increased rapidly, reaching a peak value of -16.3 \AA~ in October 
1995. This value exceeds the maximum value of -14.1 \AA~ known just before the latest $ELS$, and to our best 
knowledge, is also the highest value ever recorded. 

\begin{figure}[ht]
\vspace{0cm}
\hspace{-0.5cm} 
\psfig{file=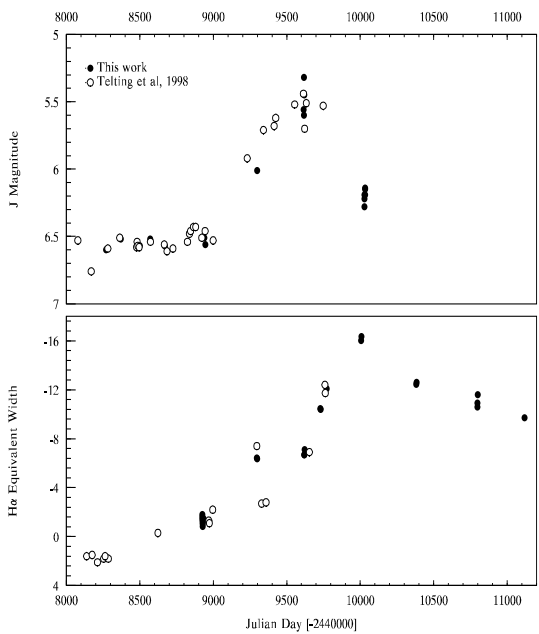, height=10cm, width=10cm, angle=0}
\caption{The equivalent width of H$\alpha$ emission line and $J$ magnitude over the course of the monitoring 
programme. H$\alpha$ $EW$ data are from Roche et al. (1993) and Kunjaya \& Hirata (1995) (open circles); 
and from this work (filled circles)}
\end{figure}

\subsection{Correlation between the infrared magnitude and the H$\alpha$ $EW$}

It is generally accepted (Neto \& Pacheco 1984; Ashok et al. 1984) that the same electrons that produce 
the excess continuum emission in the infrared are also responsible for the hydrogen emission lines which 
are a characteristic of the Be stars. In addition, the observed correlation between the H$\alpha$ and near-infrared 
fluxes of Be stars served as direct evidence of the coincidence of the IR and Balmer line emitting regions 
(Neto \& Pacheco 1984; Ashok et al. 1984; van Kerkwijk, Waters, \& Marlborough 1995).  

The correlation between the nearly simultaneous infrared $J_{ELS}$-$J$ magnitude and the H$\alpha$ $EW$ in X 
Persei is shown in Fig. 4. Some additional data are from Norton et al. (1991), Reynolds et al. (1992), and 
Roche et al. (1993). We find that a positive correlation is existence for the values at the early stage of the 
recent emission phase, consistent with the results reported by Dachs \& Wamsteker (1982) and Roche et al. 
(1993). The observations of 1994 and 1995, however, fell well off this trend. We will put forward a model 
in the next section to give a qualitative explanation for this unusual phenomenon. 

\begin{figure}[ht]
\vspace{0cm}
\hspace{-0.5cm} 
\psfig{file=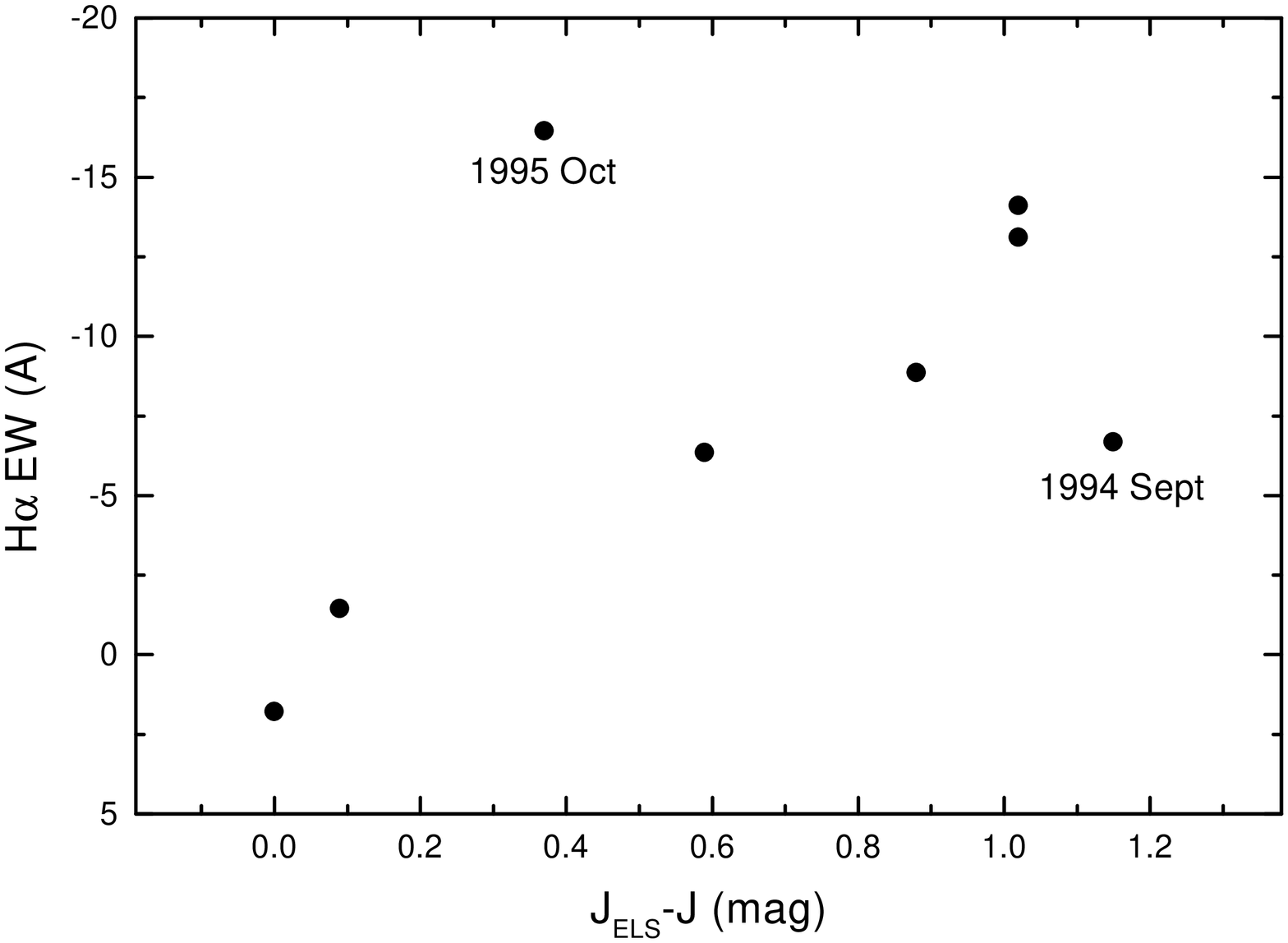, height=10cm, width=8cm, angle=-90}
\caption{The correlation between $J_{ELS}-J$ magnitude and equivalent width of the H$\alpha$ emission line in X Persei} 
\end{figure}

\section{The expanding ring model }

The observed IR excesses have been generally attributed to free-free and free-bound emission from 
envelopes around Be stars. Since the emission due to recombination declines rapidly for highly excited 
states, free-free emission dominates the IR spectrum (Chokshi \& Cohen 1988; Kastner \& Mazzali 1989). 
The free-free emission coefficient is given by
 
\begin{equation}
j_{\nu }^{ff}=6.842\times 10^{-38}N_iN_eT^{-1/2}Zg_{ff} e^{-\frac{h\nu}{kT_e}},
\label{expdec}
\end{equation}
in erg/cm$^3$/s/Hz, where $g_{ff}$ is the free-free Gaunt factor, $Z$ is the charge, $T$ is the temperature 
of the gas in $^\circ$K, $N_e$ and $N_i$ are electron and ion densities in cm$^{-3}$. The opacity is then 
obtained via Kirchoff's Law, giving

\begin{equation}
\kappa _{\nu }^{ff}=3.692\times 10^{8}N_iN_eT^{-1/2} \nu ^{-3}g_{ff} (1-e^{-\frac{h\nu}{kT_e}}),
\label{expdec}
\end{equation}
for the free-free absorption coefficient. The H$\alpha$ recombination-line intensity is given by (Tucker, 1975) 

\begin{equation}
j_{H\alpha}=1.0265\times 10^{-24}N_iN_eT^{-1/2}ln(\frac{h\nu_0}{kT}),
\label{expdec}
\end{equation}
in erg/cm$^3$/s, where $\nu _0$ is the frequency at the Lyman limit.
The absorption coefficient at the center of H$\alpha$ line is given by (Osterbrook, 1989)

\begin{equation}
\kappa _{H\alpha}=\frac{\pi e^2}{m_e c \Delta \nu _D} f_{23} N_{0,2}[1-e^ {-\frac{h\nu_{23}}{kT}}],
\label{expdec}
\end{equation}
where $f_{23}$ is the oscillator force, $N_{0,2}$ is the volume density of atoms excited in the level 2,  
$\Delta \nu _D$ is the Doppler width of the line defined by thermal motions, and the other parameters 
refer to the usual physical constants.

\subsection{The slab model}

In order to interpret the unusual variations between the near-infrared excess and the H$\alpha$ $EW$, we consider 
a model in which the equatorial matter ejected from the primary is an expanding ring with a total mass of 
$M_1$ and an initial ring width of $L_1$. We assume the ring is expanding following the polytropic law: 

\begin{equation}
TV^{n-1}=T_1V_1^{n-1}=const,
\label{expdec}
\end{equation}
where $n$ is the polytropic index, $T$ is temperature, $V$ is volume. The subscript '1' denotes the value at the 
surface of stellar photosphere ($r=1 R_*$). The temperature and density of the gas ring is $r$-related and 
decreases as the ring expanded, but at a given radius, the temperature and density are assumed to be 
uniform. We further assume that the width of the ring depends upon $r$ as $L(r) =L_1 r^ m$.

Considering a pure hydrogen envelope viewed with an angle $i$, we will calculate the luminosities from 
H$\alpha$ line emission ($L_{H\alpha}$) and from infrared continuum in the $J$ band ($L_J$) according to this 
model, by taking optical depth into account. We assume the envelope confined to a slab disc with a thickness 
$D (=2H)$. In some recent models, a cone disc with an opening angle of $\theta=5^\circ $ was taken (Marlborough, 
Zijlstra \& Waters 1997); Telting et al. 1998). However, this does not change the results fundamentally.
 
\subsection{The model parameters and results}

The double peaks at H$\alpha$ in our spectra in the strong emission phase imply that $i$ is not so small, while no 
report concerning shell lines suggests that $i$ is not close to 90$^\circ$. We adopt $i$ = 30$^\circ$ in terms of the 
derived stellar mass of 15.5 M$_\odot$, stellar radius of 7 R$_\odot$ and $vsini$~ of 200 \kms (Lyubimkov et al. 1992).  

We adopt an effective temperature $T_{eff}$ of 30000$^\circ$K, a radius $R_*$= $7R_\odot$, and  a gravitational 
acceleration $log g=4.0$ for X Persei, because they were derived during the latest disk-less phase 
(Lyubimkov et al. 1992). 
Referring to Waters et al. (1988), the density distribution for Be/X-ray binaries is given by $N(r)=N_1 r^{-n_d}$ 
with index $n_d$ between 2 and 3.75. We take $n=7/6$ and $m=2$, so that the density and temperature distributions 
are $N(r)=N_1 r^{-3}$ and $T=T_1 r^{-1/2}$ (Waters 1986) with $N_1$ and $T_1$ taken to be $10^{13}$ cm$^{-3}$ and 
30000$^\circ$K, respectively. The thickness of the slab is assumed $H=0.2 R_*$. 

With these numerical values taken, we can obtain the variations of the luminosities of the H$\alpha$ emission 
line and infrared continuum in the present model as a function of radius. The results are shown in Fig. 5 
with the luminosities normalized to their maximum values. 

Fig. 5 illustrates that the near-infrared continuum emission changes dramatically with its peak 
between $1 R_*$ and $2 R_*$. It increases to its maximum value rapidly and so does the subsequent decrease. The 
near infrared emission is almost concentrated on the regions within the radius of $2 R_*$, which is well 
consistent with the result obtained by Persi et al. (1977). The variation of the H$\alpha$ emission, however, is more 
gradual. The emission slowly increases with radius as the ring outwardly expands and dissipates. The H$\alpha$ 
emission reaches maximum much later than the near-infrared emission does. When the H$\alpha$ emission 
reaches its maximum value around radius $r=6 R_*$, the near-infrared emission is almost disappeared.  

\begin{figure}[ht]
\vspace{0cm}
\hspace{-0.45cm} 
\psfig{file=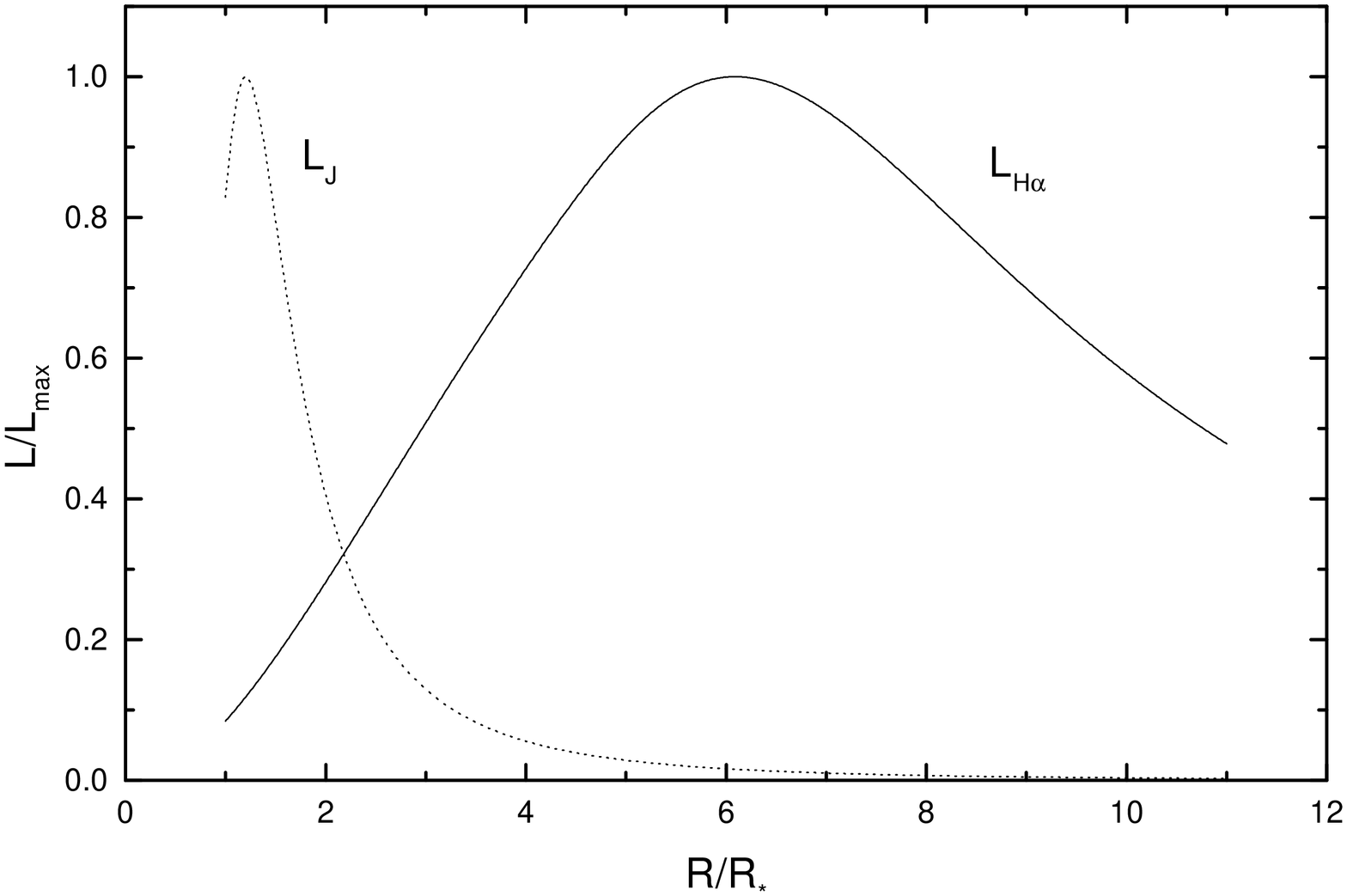, height=9.5cm, width=7.5cm, angle=-90}
\caption{The luminosity of H$\alpha$ line emission and infrared continuum emission as a function of radius, for 
the model with initial density of $10^{13}$ cm$^{-3}$ and temperature of 30000$^\circ$K. Other parameters are taken 
to be $H=0.2R_*$, $n=7/6$, $m=2$, and $i$ = 30$^\circ $}
\end{figure} 

The overall pattern (Fig. 5) is in agreement with the observational data shown in Fig. 3. However, 
we wish to emphasize here that the results obtained in this section are still preliminary. They need to be 
regarded with caution in view of the many assumptions necessary to derive them.

\section{Discussion}

Fig. 2 shows that the intensities of the H$\alpha$ emission line in 1993 and 1994 spectra did not change much 
while He I line clearly fell. This implies that the envelope has begun to dissipate. Hence, the large 
increase started in early 1995 indicats that, during 1994, before the old envelope dissipated completely, a 
new and larger one has begun to be generated. The four-peak structure in the He I profile, observed a little 
later, clearly showed the coexistence of the two envelopes.

The successive record of the H$\alpha$ activity from the beginning of the new emission phase of X Persei 
permits us to trace the evolution of the envelope in some detail. The basic pattern is as follows: at first, the 
envelope was fed with matter, and the H$\alpha$ emission intensity evolved from zero to a definite value, 
implying the beginning of the new emission phase. After having developed over a few years, by 1994, the 
envelope began to decay on a longer time-scale. The fading process, however, was interrupted by a new, 
major outburst. At this time the former contributed mainly to the H$\alpha$ $EW$ and only slightly to the $JHK$, 
while the latter gave large near-infrared excesses but contributed little to H$\alpha$, due to its proximity to the 
photosphere with a large optical depth. Thus the observed H$\alpha$ $EW$ and $JHK$ magnitudes can be produced. 
Thereafter, as the primary cease to eject matter and the ejected matter moved outwards, an envelope ring 
was formed. The outward motion of the new ring resulted in the decrease of the optical depth of the 
envelope ring due to the rapidly reduced density. When the ring reached several $R_*$, a strong H$\alpha$ and 
small infrared excess were produced (see Fig. 5). Meanwhile, the old one still generated small H$\alpha$ $EW$ 
and $JHK$ magnitudes, and so were produced the high $EW$ of H$\alpha$ and faint $JHK$ magnitudes. We suggest 
that the coexistence of the two rings may also be one of the factors for both the H$\alpha$ $EW$ and peak $JHK$ 
magnitudes exceeded their previous peak values. 

For some Be stars, the H$\alpha$ profiles did not change much over times (e.g. $\alpha$ Ara, etc.), and this may 
be related to a steady ejection of matter; for some others, the H$\alpha$ profiles underwent rapid changes (e.g. $\mu$ 
Cen; Hanuschik et al. 1993), and this may be related to occasional ejection. For steady ejection that leads 
to a steady disk-like envelope, there may exist a linear relation between the infrared excess and H$\alpha$ $EW$, 
whereas for transient ejection that leads to ring-shaped envelope, such a linear relation may not exist. We 
suggest that the variations in X Persei in recent years are intermediate: the early stage of the new emission 
phase is dominated by the early stage of a steady ejection while the later stage is likely to be related to a 
sporadic ejection.

\section{Conclusions }

We present the nearly simultaneous optical spectroscopic and near-infrared photometric observations of 
the Be star X Persei, the optical counterpart to the X-ray source 4U0352+30. Our data show that the 
$JHK$ magnitudes and H$\alpha$ $EW$ in the new phase increase to high levels that exceeded the maximum 
values ever recorded. The observed high value of the $EW$ and $JHK$ magnitudes may reflect a more extensive 
and denser envelope in the new emission phase.

The infrared photometry and the $EW$ of H$\alpha$ emission line increase coincidently in the early stage of 
the new emission phase, consistent with the results by Roche et al. (1993) for X Persei and by Dachs \& 
Wamsteker (1982) for $\mu$ Cen, $\omega$ Ori and HD58343. However, the variations for the observations in 1994 
and 1995, are largely deviated from the trend. The H$\alpha$ $EW$ hardly enhanced at the peak $JHK$ magnitudes 
in 1994, but it is seen to dramatically increase during the optical and near-infrared fade. While the latter 
got fainter dramatically, the $EW$ increased rapidly, reaching a maximum of -16.34 \AA~ in October 1995. 

We put forward a model in which the circumstellar envelope is an expanding ring, to interpret the 
unusual relation between the H$\alpha$ $EW$ and near infrared luminosities observed during the recent emission 
phase of X Persei. We suggest that the unusual phenomenon may result from the double-ring envelope, in 
which at least the new one is in the form of expanding ring. Our results indicate that although both 
infrared excess and H$\alpha$ emission line arise from the envelope, their maximum values are likely to reach at 
the different evolution stage of the envelope.

\acknowledgements We are grateful to Drs D.-M. Wei and H.-C. Wang of Purple Mountain Observatory for their valuable 
discussion and to Dr J.-Y. Wei and Mr H.-B. Li of Beijing Astronomy Observatory for their assistance in 
observations. Some data are reduced by Ms J.-P. Xia and Mr Z.-X. Zhu. We also wish to thank an anonymous 
referee for his/her useful comments and suggestions. QZL acknowledges the financial support from KC Wong
fellowship of Chinese Academy of Sciences. This work is partially supported by the Netherlands Organization for 
Scientific Research (NWO) through Spinoza Grant 08-0 to E.P.J. van den Heuvel and by the National Project for 
Fundamental Research by the Ministry of Science and Technology of China (973 project).


\begin{thebibliography}{}
\bibitem[]{}Ashok, N.M., Bhatt, H.C., Kulkarni, P.V. et al., 1984, MNRAS, 211, 471
\bibitem[]{}Braes, L.L.E. \& Miley, G.K., 1972, Nature, 235, 273 




\bibitem[]{}Chokshi, A. \& Cohen M., 1988, AJ, 94, 123 
\bibitem[]{}Corbet, R.H.D., Thomas, B., 1991, IAUC, 5372
\bibitem[]{}Dachs, J., Wamsteker, W., 1982, A\&A, 107, 579 
\bibitem[]{}Fabregat, J., Reglero, V., Coe, M.J. et al. 1992, A\&A, 259, 522 
\bibitem[]{}Hang H.-R., Xia, J.-P., 1995, Acta Astronomia Sinica, 36, 438
\bibitem[]{}Hanuschik, R.W., Dachs, J., Baudzus, M. et al, 1993, A\&A, 274, 356
\bibitem[]{}Huang, S.-S. , 1972, ApJ, 171, 549
\bibitem[]{}Huchings, J.B., 1977, MNRAS, 93, 486 
\bibitem[]{}Kastner, J.H. \& Mazzali, P.A., 1989, A\&A, 210,295
\bibitem[]{}van Kerkwijk, M.H., Waters, L.B.F.M., \& Marlborough, J.M., 1995, A\&A, 300, 259
\bibitem[]{}Kunjaya, C., Hirata, R., 1995, PASJ, 47, 589 
\bibitem[]{}Liu Q.-Z., Hang H.-R., 1997, Acta Astronomia Sinica, 38, 434
\bibitem[]{}Lyubimkov, L.S., Rostopchin, S.I., Roche, P., Tarasov, A.E., 1997, MNRAS, 286, 549
\bibitem[]{}Marlborough, J.M., Zijlstra, J.W., \& Waters, L.B.F.M., \& 1997, A\&A, 321, 867
\bibitem[]{}Mook, D.E., Boley, F.I., Foltz, C.B. et al., 1974, PASP, 86, 894 
\bibitem[]{}Neto, A.D. \& Pacheco, J.A. de Freitas, 1982, MNRAS, 198,659 
\bibitem[]{}Norton, A.J., Coe, M.J., Estela, A. et al., 1991, MNRAS, 253, 579 
\bibitem[]{}Osterbrock, D.E, 1989, Astrophysics of Gaseous Nebulae and Active Galactic Nuclei, University Science 
Books, Mill Valley
\bibitem[]{}Penrod, G.D. \& Vogt, S.S., 1985, ApJ, 299, 653
\bibitem[]{}Persi, P., Viotti, R., Ferrari-Toniolo, M., 1977, MNRAS, 181, 685 
\bibitem[]{}Reynolds, A.P., Hilditch, R.W., Bell, S.A..et al, 1992, MNRAS, 258, 439 
\bibitem[]{}Roche, P., Coe, M.J., Fabregat, J. et al. 1993, A\&A, 270, 122
\bibitem[]{}Roche, P., Larionov, V., Tarasov, A.E. et al. 1997, A\&A, 322, 139
\bibitem[]{}Tarasov, A.E. \& Roche, P., 1995, MNRAS, 276, L19
\bibitem[]{}Telting, J.H., Waters, L.B.F.M., Roche, P. et al, 1998, MNRAS, 296, 785
\bibitem[]{}Tucker, W.H., 1975, Radiation processes in astrophysics, MIT Press.
\bibitem[]{}Waters, L.B.F.M., 1986, A\&A, 162, 121
\bibitem[]{}Waters, L.B.F.M., Taylor, A.R., van den Heuvel, E.P.J. et al., 1988, A\&A, 198, 200
\bibitem[]{}Zamanov, R.K. \& Zamanova, V.I., 1995, IBVS, No. 4189 

\end{thebibliography}
\end{document}